\documentclass[aps, prd, showpacs, superscriptaddress, groupedaddress, twocolumn]{revtex4} 

\makeatletter
\def\@society{generic}
\makeatother
\usepackage{titlesec}
\usepackage[a4paper, margin=0.6in]{geometry}
\usepackage{graphicx}  
\usepackage{subcaption}
\usepackage{float}
\usepackage{tabularx}
\renewcommand{\thesection}{\arabic{section}}
\usepackage{array}
\usepackage{amssymb}
\usepackage{mathtools}
\usepackage{dcolumn}
\usepackage{color}
\usepackage[font=small,justification=raggedright]{caption}
\titleformat{\section}[hang]
  {\normalfont\large\bfseries}{\thesection}{1em}{}
\usepackage[pagebackref=false, colorlinks=true]{hyperref}
\hypersetup{
linkcolor=blue,
citecolor=blue, 
urlcolor=blue}   
\usepackage{amsmath}
\usepackage{ulem}
\usepackage[utf8]{inputenc}

\def\be {\begin{equation}}
\def\ee {\end{equation}}
\def\ba {\begin{eqnarray}}
\def\ea {\end{eqnarray}}
\usepackage{caption}
\newcommand{\barray}{\begin{array}}
\newcommand{\earray}{\end{array}}
\usepackage{soul}
\usepackage{amsmath}

\begin{document}
\title{On Formation of Primordial Naked Singularities}
\author{Koushiki}
\email{koushiki.malda@gmail.com}
\affiliation{International Centre for Space and Cosmology, School of Arts and Sciences, Ahmedabad University, Ahmedabad, GUJ 380009, India}
\author{Pankaj S. Joshi}
\email{psjcosmos@gmail.com}
\affiliation{International Centre for Space and Cosmology, School of Arts and Sciences, Ahmedabad University, Ahmedabad, GUJ 380009, India}
\author{Sudip Bhattacharyya}
\email{sudip@tifr.res.in}
\affiliation{Department of Astronomy and Astrophysics, Tata Institute of Fundamental Research, Homi Bhabha Road, Colaba, Mumbai 400005,
India}

\date{\today}

\begin{abstract}
The density fluctuations in the nearly homogeneous background in the very early universe are argued to be the origin of the cosmic structures we observe in our present universe. Along with many other structures, these fluctuations would have also given rise to primordial black holes at the end of unhindered gravitational collapse of high-density matter blobs that developed due to these fluctuations. We study here such a collapse, \textcolor{black}{which are seeded by a scalar field $\phi$ associated to a non-trivial potential function $V(\phi)$, minimally coupled to gravity. Such a continual collapse is presumed to form a black hole always and is named a primordial black hole (PBH).} Examining the dynamics of such a collapse, we find the parameter range where the apparent horizon does not form, thus resulting in the visibility of the final singularity of collapse for faraway external observers. This treatment is within the classical limits dictated by Planck's constraints. The slow-roll parameters are analysed here to keep the relic abundance of the scalar field high enough so that the abundance of produced primordial naked singularities (PNaSs) falls within the range of resolution of possible observational probes.
\\\\
$\textbf{key words}$: Unhindered gravitational collapse (UGC), Apparent Horizon (AH), Primordial black hole (PBH), Primordial naked singularity (PNaS).

\end{abstract}
\maketitle

\section{Introduction}\label{intro}
Primordial universe is intriguing and also an area of active research in various sub-fields in physics, including particle physics, gravitational physics, astrophysics, and cosmology. The reason behind the formation of all objects, ranging from supermassive black holes to dark matter particles and the dynamical properties of the current universe, from its accelerated growth to its homogeneous and inhomogeneous structures in all length-scales, is postulated to have been due to the dynamics in the very early universe. 
From CMB \cite{CMB} and BAO \cite{BAO} observations, it has been observed that the early universe is typically homogeneous on large enough scales. Although, these very observations also confirmed that there are  small fluctuations in this overall homogeneous background.

Naturally, there is tremendous scientific interest in studying these fluctuations, whose origin lies in quantum fluctuations from the very early universe, which affect the structures and dynamics of both the early and later universe. Zel'dovich and Novikov, in $1967$, postulated that these density fluctuations could lead to gravitational collapse to form black holes in the very early universe, which later became known as primordial black holes (PBHs). 
In $1971$, Hawking postulated that these fluctuations gave rise to collapse in very small regions of space and produced black holes of mass as small as $10^{-5}g$ \cite{Hawking1971}.
These works, along with Carr and Hawking's work \cite{Carr1974}, three years later, established the relevance of probing the nature of PBHs. 
After this, PBH has been an area of active research. The epoch of formation and mass spectrum of PBHs \cite{Carr1975, Zel'dovich1967, Sendouda2003} have been the topic of much active debate. This is also because of the current availability of huge observational datasets, which can be used to define constraints on the masses of PBHs \cite{Carr2010, Kanori}. These observations \cite{Afzal2024}, mainly of gravitational waves, microlensing constraints on PBH masses etc. can help in narrowing down the large class of inflationary potentials for early universe \cite{Linde1983, Mukhanov1985, Lucchin1985, Albrecht1982, Linde1994, Freese1990, Starobinsky1980, Bezrukov2008}, based on which the masses of the PBHs are thought to be decided. The perturbations generated by these potentials are said to trigger the collapse of blobs of matter which in turn formed PBHs during and after the inflationary era, if the abundance of these fields were high enough. 

There is an array of work on the formation of PBHs in the early universe epoch. Curvature perturbation to trigger collapse and form PBHs have been discussed in \cite{Ozsoy2023, Frolovsky:2011FKS},  while \cite{Tada2023, Geller2022} discuss how multifield-potentials, minimally or non-minimally coupled to gravity, trigger perturbations high enough to produce PBHs. Karam et.al.\cite{Karam2023} discussed the effect of double well potentials to generate such perturbations and Solbi et.al.\cite{Solbi2021} did a similar analysis involving quartic potentials. While the most discussed end-state of primordial gravitational collapse has been PBHs, there are some non-trivial results where the end-states have other causal characters \cite{Pi:2017gih}. In these works, the prime foci have been estimating and constraining the fluctuation $\delta \phi$ in the inflationary scalar field $\phi = \bar{\phi} (t) + \delta \phi (r,t)$ and observe whether the fluctuations generated from a specific potential can trigger collapse to form a PBH. Khlopov \cite{Khlopov2010} and Harada \cite{Harada2016} have also analysed the amount of perturbation and calculated the abundance of PBHs and their masses. \textcolor{black}{It is important to mention here that PBHs can form just after Planck time \cite{Carr2021} with a mass as low as $10^{-5}$g and this is in fact what Hawking \cite{Hawking1971} proposed, but the formation process has to be studied with much nuance, if the formation is triggered by inflationary potentials. Though the formation of PBHs is possible even during inflation but, the number density of PBHs produced during inflation would be diluted beyond the limit of observational capacities \cite{carr2025primordialblackholes}. The very rapidly accelerated expansion of this epoch will push the freshly formed PBHs so far from each other that they will be observationally insignificant with current probes, even if they do not evaporate. But the effects of inflation might live beyond this phase and seep into the epochs of reheating and radiation domination.} 

What is important to note here is, a crucial aspect is yet to be addressed in this methodology and the body of work, which is the dynamical evolution of the collapse of  perturbed collapsing clouds. Joshi and Dwivedi showed the importance of analysing the dynamics of  gravitational collapse of perfect fluids and other matter fields with both zero \cite{Joshi1991} and non-zero pressures \cite{Joshi:1994st} and the necessity of this methodology in determining the visibility or otherwise of the singularity that forms at the end of the collapse. Later, it was also shown that various minimally coupled scalar fields can also lead to singularities, which are visible \cite{Goswami:2007, Mosani2022}. Also, depending on the equation of states corresponding to these scalar fields, these were classified into the ones that produce black holes and naked singularities at the end of a gravitational collapse \cite{KK1}. Recently, Broderick \cite{Broderick2024}, while suggesting from an observational perspective, stated that the galactic centers,  $Sgr A*$ and $M87*$, may not harbor a naked singularity. However, this work pointed out that the gravitational collapse  solutions giving rise to naked singularities are nevertheless generic. Their study is based on the trajectories of particles in-falling towards the naked singularity and the observational signatures given out by the same ones. And they have pointed towards its limitation of not capturing the possibility that such singularities could emit non-spacelike geodesics that could potentially alter the structure of space-time and make the singularity observable to distant observers as well. \textcolor{black}{Moreover, the stability analysis of homogeneous perfect fluid clouds has shown that perturbation in the initial data might change the final outcome from a black-hole (BH) to a Naked Singularity (NaS) \cite{Satin:2014ima}, thus making NaSs more stable. This has been reaffirmed for Vaidya NaS as well \cite{Wheeler:2022xdo}.}

Recently, it has been proposed and pointed out for the first time that primordial fluctuations could have given rise to visible or naked singularities
\cite{JoshiBhattacharyya2025}.
If confirmed, such primordial naked singularities could possibly change the current view about the contents of the universe. 
For example, a significant fraction of the dark matter, and hence the current universe could be made of visible singularities.
Such a large abundance of PNaSs not only addresses the physics of dark matter but also provides potential natural laboratories to probe the effects of quantum gravity and to test the corresponding theories. 
However, this work does not delve into the intricacies of the dynamics of the formation of PNaSs. 
The novelty and purpose of our current work lies in addressing this important aspect.
\textcolor{black}{The Early universe is a hot soup of high-energy particle physics, and there are multiple scalar-field models to explain these interactions in particle cosmology. For instance, the theory of chaotic inflation is based on polynomial scalar-field potential \cite{Linde1983}. Massive scalar fields also are associated with a polynomial potential \cite{Mukhanov1985}. Polynomial functions are also used in more complicated scalar-field potentials \cite{Lucchin1985, Albrecht1982, Freese1990}. Whereas, there are works that attempt to quantise such scalar field perturbations. Some of them use massless scalar fields \cite{Starobinsky1980}. Higgs potential is another form of potential \cite{Bezrukov2008} that is used in such studies that consider non-minimal interaction with the gravitational field. Multiple such models use exponential potentials as well \cite{Frolovsky:2011FKS}. Other than these, there are multi-field potentials to describe these interactions \cite{Linde1994}, but for the present purpose, we do not go into these. So, while considering the gravitational collapse of density fluctuations in early universe, following a scalar field with an arbitrary non-trivial potential $V(\phi)$ dynamically will give us more insights into the visibility of the resultant singularity}, thus informing us on the formation of PBHs or PNaSs, while maintaining the generality of the class of scalar field potentials. So, in our work, the perturbations are triggered by a general potential, minimally coupled to gravity. We also discuss the constraints on this potential so that the methodology is still classical and treatable within a theory of perturbation. We also discuss the constraints on the slow-roll parameters required to maintain a sufficiently high relic abundance of this potential, ensuring that the production of PNaS is observationally relevant. \textcolor{black}{The theoretical study of PBH formation often ignores the nuance of the physical process of the overdense bubble collapsing into a singularity while the sparser background keeps expanding. In different evolutionary epochs, the expansion is even accelerated. We discuss this through generalized matching conditions.}

\section{Dynamical constraints on a homogeneous scalar field collapse}\label{FS}
Primordial singularities can form in different evolutionary epochs of the universe. Depending on the epoch of their formation their mass can widely vary, but there are prohibited epochs where these singularities are not formed. Since we are interested here mainly in examining the collapse of scalar fields to form such singularities, we are restricting our focus to the inflationary era when the evolution is dominated by scalar fields with associated non-zero potentials. In this era, the universe is governed by a flat FLRW metric with line element:
\be\label{flrw}
ds^2 = -dt^2 + a^2 dr^2 +{R(a)}^2 \; d\Omega^2\;,
\ee
where, $R(a)= r a(t)\equiv$ physical radius of the collapsing cloud, $r\equiv$ co-moving radius that stays unchanged for an observer whose rest frame coincides with the frame of collapsing cloud, $a(t)\equiv$ scale-factor that changes in the range $[1,0)$ and $d\Omega^2\equiv$ line element of the unit 2-sphere. In this manifold $\mathcal{M}$, the homogeneous scalar field: 
\begin{equation}
    \phi: \mathcal{M} \to \mathbb{R}\;,
\end{equation}
becomes $\phi\equiv \phi (a)$. We have seen in \cite{KK1, DeyK:2023, Koushiki:2024rwt} that the potential function $V(\phi)$ of the scalar field plays a crucial role in deciding the final fate of the collapse. As we want to investigate the  non-trivial causal structure of the end-state singularity, we only consider $V(\phi)\neq 0$. For this case, the Lagrangian is given by:
\begin{equation}\label{lagra}
    \mathcal{L}_{\phi}=-\frac{1}{2}g^{\mu \nu}\partial_{\mu}\phi \partial_{\nu} \phi-V(\phi)\;.
\end{equation}
Varying the action associated to the Lagrangian with respect to $g_{\mu\nu}$, we get the stress-energy tensor for the scalar field: 
\be\label{stress1}
T_{\mu\nu}=-\frac{2}{\sqrt{-g}}\frac{\delta \left(\sqrt{-g} \mathcal{L}_{\phi}\right)}{\delta g^{\mu\nu}}.
\ee
Now, we can write down the equations of motion for the scalar field using Eqs.(\ref{flrw}, \ref{lagra}, \ref{stress1}):
\ba 
\rho &=& \frac{1}{2}\dot{\phi}^2 +V \label{rho}= \frac{3\dot a^2}{a^2}\\ \label{p}
    p&=&\frac{1}{2}\dot{\phi}^2-V=-\frac{2\ddot a}{a}-\frac{\dot a^2}{a^2}\;,
\ea
where, $``\cdot"$ denotes the partial derivative with respect to the co-moving time. There is another equation of motion for the scalar field, the Klein-Gordon equation:
\be \label{KG}
\ddot{\phi}+ 3\frac{\dot{a}}{a} \dot{\phi} + \frac{\partial }{\partial \phi}V(\phi)=0\;,
\ee
which is just the equation for the conservation of energy $\partial_\mu T^{\mu \nu}$ using Eqs.(\ref{lagra}, \ref{stress1}).
We assume a proportionality between the density and the pressure of the fluid:
\be\label{eos}
p= \omega (a) \rho\;,
\ee
and it is free to change as the scale-factor changes during the collapse. The DESI collaboration recently pointed towards the possibility of $\omega$ being a function of space-time co-ordinates \cite{desicollaboration2024}. The special cases of $\omega$ being a constant are specific examples of this larger class.

\textcolor{black}{At present scalar fields are postulated to cause the accelerated expansion of the universe only, whereas in the early universe, the scalar fields might have worked as mediators of fundamental forces and played a key role in gravitational processes as well. This is valid if the standard model theory of fundamental forces are true and at very early times, all these forces are equivalent. The potentials used to describe these interactions are polynomials, exponential or trigonometric \cite{Linde1983, Mukhanov1985, Lucchin1985, Albrecht1982, Linde1994, Freese1990, Starobinsky1980, Bezrukov2008}. As discussed before, we consider only single field potentials and minimal interaction with gravity. We begin our analysis with a general potential with which the potential function,
\be \label{V}
V(\phi) \equiv V(\phi (a)) \equiv V(a)\;.
\ee 
We keep the potential arbitrary to include all the potential forms, mentioned above.} For such scalar fields associated to non-zero potential minimally coupled to gravity, to have enough relic abundance post inflation, into the reheating era and provide the overdensity to form primordial singularities, the slow roll parameters must behave in a certain way and we must have \cite{Carr2022, Carr1993, Thoss2024}:
\be \label{Vprime}
\frac{V''}{V}\; > \frac12 {\left(\frac{V'}{V}\right)}^2\;.
\ee 
\textcolor{black}{This constraint is re-established through recent observations \cite{Carr2010, Green:2020jor} as well.} This effectively means that larger co-moving wave-number $(k=aH)$ has larger amplitude of inflationary fluctuations. The change of the amplitude of the fluctuation with respect to the mass of the primordial singularity has to be less than zero, and for that to occur Eq.(\ref{Vprime}) is necessary. In summary, the smaller mass-scales or larger $k$ is associated with larger amplitude fluctuation and the spectrum is blue-tilted. This is a non-trivial behaviour for the inflationary fluctuations. Other than the mechanism explained in Eq.(\ref{Vprime}), it can also be explained by many other mechanisms \cite{Wang:2024vfv, Pi:2017gih}.
So, Eqs.(\ref{rho}, \ref{p}, \ref{KG}) are the governing equations for the collapse of the scalar field. We are concerned with perfect fluids only, as the early universe is comprised of a perfect fluid, with isotropic pressure. Also, Eq.(\ref{Vprime}) is another governing equation, which establishes the domination of curvature over the slope of the potential. We keep the pressure a function of the scale factor and this will also decide the visibility or otherwise of the end-state singularity. We also use Eq.(\ref{Vprime}) to put bounds on the parameter of the potential function. These assumptions are realistic as $\phi=0$ is not possible, because then the scalar field vanishes identically. Also, $\dot{\phi}\neq0$ as otherwise the fluid under consideration becomes stiff, which is almost never true in the very early universe we are considering. \textcolor{black}{ These potential forms are constrained to \cite{Book1}:
 \ba 
 \partial_i\phi \partial^i\phi &\lesssim& M_P^4\label{partphi},\\
 V(\phi) &\lesssim& M_P^4\label{vphiineq},\\
 R_{ijkl}R^{ijkl}&\lesssim& M_P^4\label{Rcons},
 \ea }
\textcolor{black}{where, $M_P\equiv \sqrt{\frac{\hbar c}{ G}}$ is the Planck length} The constraint Eq.(\ref{partphi}) signifies that the effect of this potential is perturbative and does not exceed the Planck scale. $V(\phi)$ is the potential energy of the scalar field and it must not exceed the Planck scale. Eq.(\ref{vphiineq}) ensures this. Eq.(\ref{Rcons}) ensures that the deformation of the space-time due to this potential does not exceed Planck energy. Otherwise, this investigation would need the intervention of a theory of quantum gravity, which we do not currently have. \textcolor{black}{Our calculations are consistent with these constraints for a general form of $V(\phi)$, although we leave the exact values of the constants associated to each of these potentials to be restricted case by case.}

\section{Unhindered collapse of a homogeneous scalar field}\label{UGC}

Gravitational collapse occurs when the time-like congruence representing the particle or matter trajectories is focused to a point in the spacetime manifold, making the expansion scalar of the time-like congruence negative: $\Theta=\nabla_a u^a < 0$. This means that $\dot{R} < 0$. Also, $r$ is the co-moving radius and a strictly non-negative quantity that stays constant over time for a co-moving trajectory. So, in this case the scale-factor only changes and we have $\dot{a} < 0$. The collapse is unhindered if there is no obstacle to stop its evolution and the end-state is at $\Theta\to -\infty$ and $R= a\to 0$. Evidently, there is a singular state at the end of such a collapse and the singularity is the boundary of the manifold. At this point, we are only allowed to use limiting values of the functions and not the absolute ones. 

From Eq.(\ref{rho}), we get the value of $\dot{a}$:
\begin{equation}\label{adot}
    \dot a=-\sqrt{\frac{\rho(a)}{3}}a,
\end{equation}  
and we keep only the $-ve$ root as per the preceding discussion. Now, adding Eqs.(\ref{rho}, \ref{p}), we get:
\begin{equation}\label{rhop}
\rho+p=\phi_{,a}^2\dot a^2\;,
\end{equation}
where, we use chain rule to write $\dot{\phi}= \frac{\partial \phi}{\partial a} \frac{\partial a}{\partial t}= \phi_{,a} \dot{a}$. Differentiating Eq.(\ref{adot}) we get:
\begin{equation}\label{ddota}
    \ddot a =\frac{1}{3}a\left(\frac{a \rho_{,a}}{2}+\rho\right).
\end{equation}
Using Eq.(\ref{rho}) and (\ref{p}), with replacing $\dot{a}$ and $\ddot{a}$ using Eq.(\ref{adot}) and (\ref{ddota}), we have obtained the density as a function of $a$ as
\begin{equation}\label{rho(a)}
    \rho=\rho_0 \exp{\left(\int_a^1 a \phi_{,a}^2 da\right)}.
\end{equation}
This is one key equation to follow the \textit{Unhindered Gravitational Collapse} (UGC), because in the neighbourhood of the singularity, as $a\to 0$, density, as well as pressure blow up. Using Eq.(\ref{adot}) in Eq.(\ref{rhop}), we get
\begin{equation}\label{rhop2}
    \rho\left(1-\frac{\phi_{,a}^2a^2}{3}\right)+p=0.
\end{equation}
To express the potential in terms of the dynamical quantities, we add Eqs.(\ref{rho}, \ref{p}) and replace in Eq.(\ref{rhop2}) to arrive at:
\begin{equation}\label{rhofinal}
    V(\phi)=\rho\left(1-\frac{\phi_{,a}^2a^2}{6}\right).
\end{equation}

Using Eqs.(\ref{adot}, \ref{rhop}) in Eq.(\ref{p}), we now obtain:
\begin{equation}\label{rhoabyrho}
    \frac{\rho_{,a}}{\rho}=-\frac{\phi,_a^2}{a}.
 \end{equation}
Now, using Eqs.(\ref{eos}, \ref{rhop}), we get the required relation between the potential function and the equation of state:
 \begin{equation}\label{vrhoomega}
    V(\phi)=\frac{\rho}{2}\left(1-\omega (a) \right).
\end{equation}
Equating the left hand sides of Eqs.(\ref{rhofinal}, \ref{vrhoomega}), we have
\begin{equation}\label{phiaomega}
    \phi(a),_{a}=\pm \frac{\sqrt{3\left(1+\omega(a)\right)}}{a}.
\end{equation}
In the course of the evolution, the scale factor changes in the range $a\in (0,1]$ and this defines our domain of integration. As $a=0$ is never reached in the manifold structure and it has to be observed as a limiting point, we keep the lower limit undefined and arrive at: 
\begin{equation}\label{phiomega}
    \phi(a)=\pm \int_a^1\frac{\sqrt{3\left(1+\omega(a)\right)}}{a}da+c.
\end{equation}
We will, later, take $a\to 0$. Replacing Eq.(\ref{rho(a)}) in Eq.(\ref{rhofinal}) we get:
\begin{equation}\label{vomega}
    V(a)=\rho_0 \left(\frac{1-\omega(a)}{2}\right)\exp\left( \int^1_a \frac{3\left(1+\omega(a)\right)}{a}da\right),
\end{equation}
where, the equation of state decides the potential function. To calculate the potential function in terms of the scalar field and its derivatives, we can also get from Eqs.(\ref{rho(a)}, \ref{rhofinal}):
\be\label{Vphi}
V(\phi) =\rho_0 \exp{\left(\int_a^1 a \phi_{,a}^2 da\right)} \left(1-\frac{\phi_{,a}^2a^2}{6}\right).
\ee 
This equation can also be written in terms of the scalar field and its potential:
\be\label{pde1}
\frac{V_{,\phi}}{V} +\frac{a}{3} \frac{a\phi_{,aa}+\phi_{,a}}{1-\frac{a^2\phi^2_{,a}}{6}}+a\phi_{,a}=0\;,
\ee
which can be solved for different potential functions using Eq.(\ref{V}) for different values of $n$. 
We will now first look at the criterion of visibility and then use this equation for obtaining specific solutions.

\section{Visibility of the end-state singularity}
A singularity, formed as the end-state of a UGC is termed naked if there exists a family of causal geodesics whose past lies in any arbitrary close neighbourhood of the singularity. If the future of such geodesics is complete which go away to faraway observers then the singularity is said to be globally naked. If it is incomplete then the singularity is locally naked. Before identifying a globally Naked Singularity (NaS), we first look at local visibility, which is denoted by the expansion scalar of the outgoing null congruence,
\be\label{thetal}
\Theta_l = 1-\frac{F}{R}\;,
\ee
where, $F$ is the mass enclosed within a comoving radius $r$ at co-ordinate time $t$, commonly known as Misner-Sharp mass \cite{Misner:1964}. This is given by,
\be\label{ms}
F=\int_0^R \rho(a) dr = \frac13 \rho(a) R^3.
\ee
$\Theta_l<0$ denotes trapped surfaces, $\Theta_l>0$ denotes non-trapped surfaces and $\Theta_l=0$ is the boundary between these two cases and denotes the \textit{Apparent Horizon} (AH). If trapped surfaces form and there is an AH, then if it is outgoing causal, then the singularity is locally naked.

\begin{figure}[ht!]
    \centering
    \begin{subfigure}[b]{0.4\textwidth}
        \centering
\includegraphics[width=\textwidth]{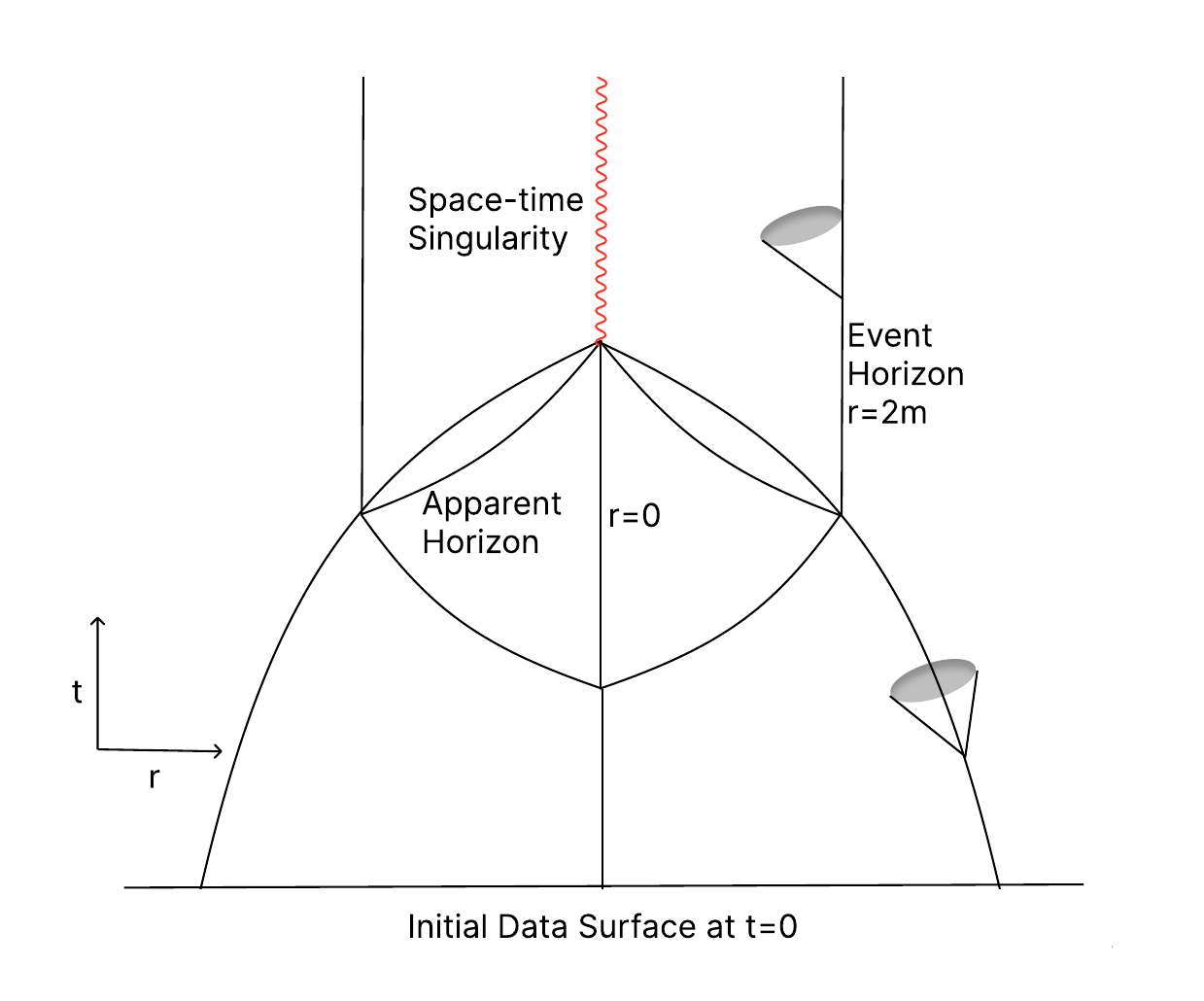} 
        \caption{Formation of BHs.}
        \label{sub1}
    \end{subfigure}
    \caption{In the cases where a BH forms as an end-state, the event and the apparent horizons form before the formation of the singularity. Hence, there are no outgoing radial null geodesics (ORNGs) and the singularity is not visible from outside.}
    \hspace{15mm}
\end{figure}
\begin{figure}[ht!]
    \begin{subfigure}[b]{0.4\textwidth}
        \centering
\includegraphics[width=\textwidth]{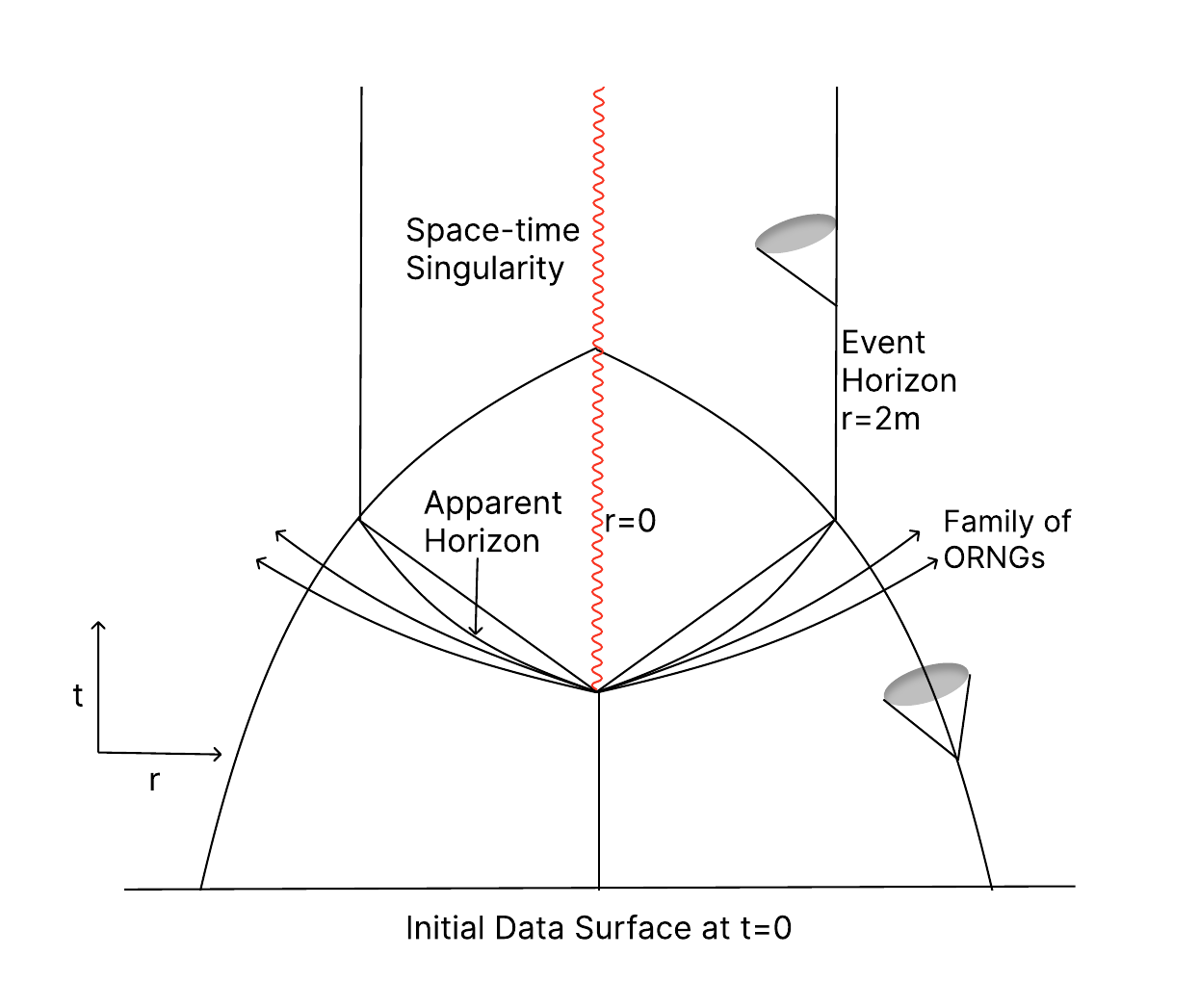} 
        \caption{Formation of globally event-like NaSs.}
        \label{sub2}
    \end{subfigure}
\caption{In this formation scenario, the time of formation of the singularity and the event and apparent horizons are the same. Hence, there is a family of outgoing radial null geodesics in the spacetime and the singularity can be seen by outside observers.}
    \label{fig1}
\end{figure}
\begin{figure}[ht!]
    \centering
    \begin{subfigure}[b]{0.4\textwidth}
        \centering
        \includegraphics[width=\textwidth]{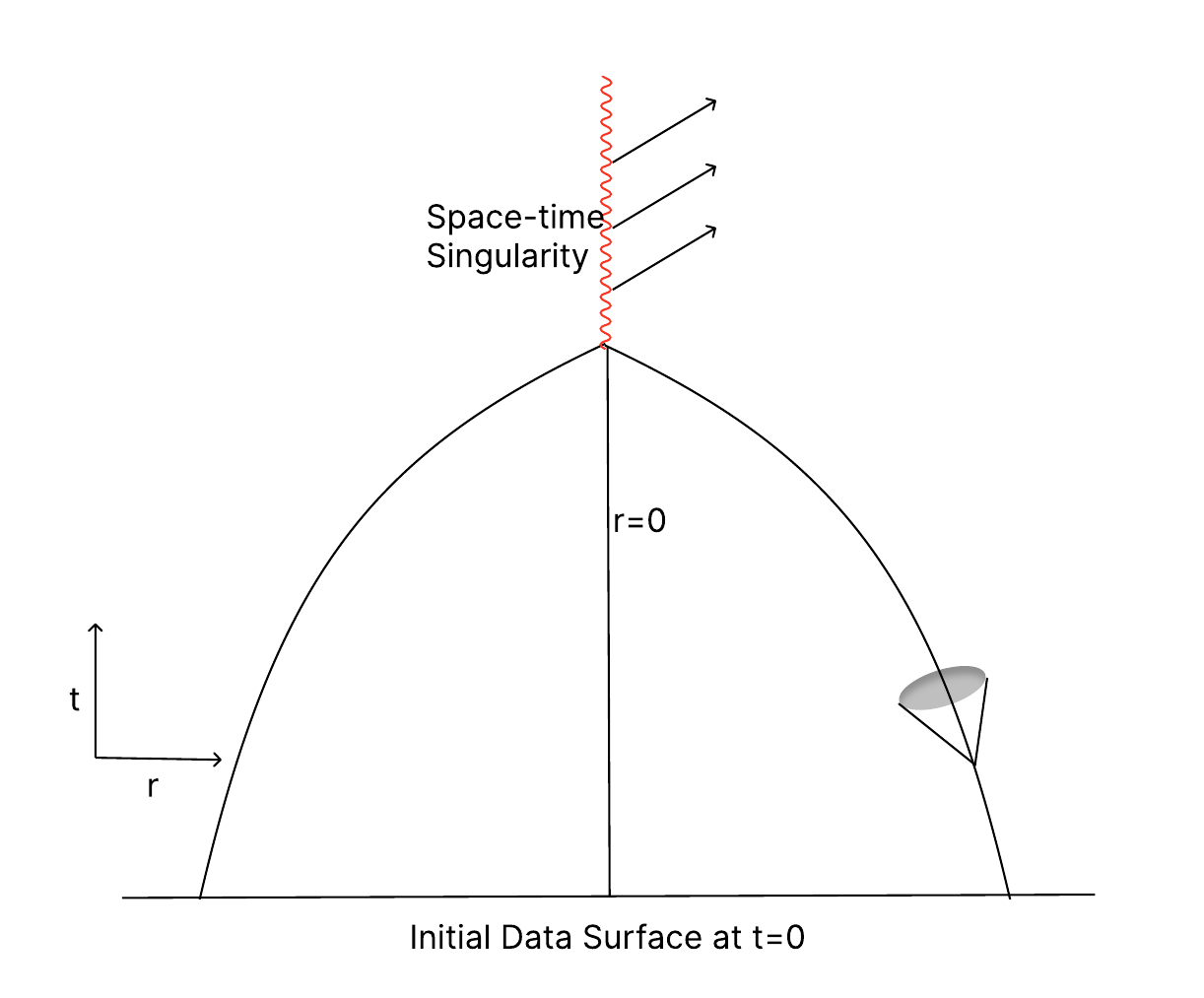} 
        \caption{Formation of globally object-like PNaSs.}
        \label{sub44}
    \end{subfigure}
\caption{The NaS at the end-state of a gravitational collapse is globally naked in this case. There are no trapped surfaces and there will be causal null geodesics emerging from the singularity, which are future complete. The singularity formed in this situation does not form any horizon (either EH or AH) around it ever.}
\end{figure}

The visibility of a singularity is determined by the existence of a family of outgoing radial null geodesics (ORNGs). These two figures clearly depict the necessity to analyse gravitational collapse dynamically. In Subfig.(\ref{sub1}) the singularity develops simultaneously for the whole collapsing cloud and entirely covered within the event horizon. But in Subfig.(\ref{sub2}), the cloud does not collapse at once to the central singularity. It can be seen clearly as the singularity curve (red zigzag line) starts forming before the whole cloud collapses. At the instant when the central singularity forms, AH also forms. This leads to outgoing causal geodesics coming out of the singularity. This singularity subsequently gets hidden within the horizon. If there is no formation of an AH, the singularity is globally visible. The necessary and sufficient condition for such a scenario is, using Eq.(\ref{thetal}, \ref{ms}),
\be\label{AH}
\lim_{a\to 0} \frac{\rho (a) r_b a^2}{3}<1.
\ee 
For all such configurations, where this inequality holds, EH or AH never forms and the singularity is visible to all distant observers at all later epochs. Thus, this singularity is globally naked and object-like. A schematic diagram of its formation is shown in Fig.(\ref{sub44})
\vspace{-0.5cm}
\textcolor{black}{
\section{Some specific examples}
There are multiple potential forms, motivated by high energy particle physics, which are proposed to dominate in the early universe. Below, we discuss some specific examples for which there is a range of allowed parameter values, where the end-state of an unhindered gravitational collapse is a PNaS. 
\subsection*{Polynomial potential}
One of the potential forms that is a special case of $V(\phi)$ is a polynomial potential. This form of potential represents chaotic inflation, first hypothesized by Linde \cite{Linde1983, Carr1993}:
 \be 
 V(\phi) = \frac{\lambda \phi^n}{n{(M_P)}^{n-4}},
 \ee
 and $n>0$ and $0<\lambda<1$ are scaling constants. The other forms of potentials also have constants, which are crucial to understand the scales of physical interactions. But, in the present analysis, we have gotten rid of such constants for the ease of calculations. But, they can be easily re-integrated into these calculations to give a clear idea about the proper length scales for each of such cases. Now, substituting Eq.(\ref{V}) in Eq.(\ref{Vprime}), we get the necessary and sufficient condition for the scalar field to accelerate enough:
\be \label{accV}
n> 2\;,
 \ee 
with the assumptions $\phi\neq n\neq 0$. Now, subtracting Eq.(\ref{p}) from Eq.(\ref{rho}) and using Eq.(\ref{V}, \ref{eos}) in Eq.(\ref{AH}), we arrive at:
\be \label{global}
\lim_{a\to 0} \frac{\phi^n}{1-\omega (a)}< \lim_{a\to 0} \frac{1}{a^2}\;,
\ee 
and this is with the assumption that the boundary of the cloud $r_b>0$.}
\begin{figure}[ht!]
        \centering
    \begin{subfigure}[b]{0.4\textwidth}
\includegraphics[width=\textwidth]{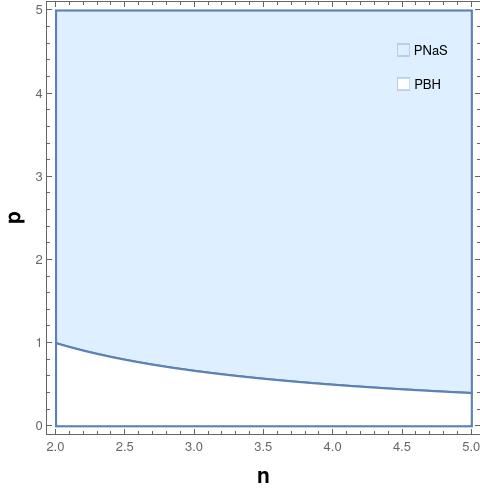}
    \caption{Parametric regions allowing the formation of PNaSs (sky blue) and PBHs (unshaded).}
        \label{param}
    \end{subfigure}
    \hspace{25mm}
\caption{The inflationary potential $V(\phi)=\phi^n$ and the scale factor vary with the scalar field as $\phi(a)\propto 1/{a^p}$. This figure depicts the interplay between these two parameters $n$ and $p$. In this parametrically allowed space (shaded region), the schematic diagram of the PNaS formation looks like Fig.(\ref{sub44}).}
    \label{fig2}
\end{figure}
\textcolor{black}{We have seen that for dust in a flat collapsing FLRW space-time \cite{KK1}, the end-state is black-hole. Carr and Kuhnel in \cite{Carr2022} restrict the equation of state to be $\omega(a)< 1$. If $\omega(a)< -\frac13$, then the strong energy condition is violated and Eqs.(\ref{rho}, \ref{p}) show that $\ddot{a}>0$ in this case. In such a scenario, the collapse might be halted and the end-state might not be singular. This falls outside the extent of our current discussion and we refrain from commenting on the same. With $-\frac13<\omega<1$, the denominator of the left hand side of Eq.(\ref{global}) is always $+ve$. So, we are left with the inequality:
\be \label{global1}
\lim_{a\to 0} \phi^n< \lim_{a\to 0} \frac{1}{a^2}\;.
\ee 
For realistic scenarios, $\phi (a) $ should increase as $a$ decreases, and hence a natural choice is $\phi(a)\propto \frac{1}{a}$. We are restricted to $n> 2 $ by Eq.(\ref{accV}). Hence, for such a scalar field, the end-state singularity is a black-hole. 
For a more general scenario, $\phi(a)\propto \frac{1}{a^p}$ for all values $-np<2$, produces a globally naked singularity, as there is no AH formation. Fig.(\ref{param}) shows the relative parametric dependence of the scale factor and the inflationary potential, responsible for the visibility of the end-state singularity of the collapse. It is easily seen from this representative figure, that for a very generalised scenario, the probability of occurrence of NaSs  is higher than that of the BHs. Also, the occurence of BHs are not ruled out, just the possibility of the occurrence of NaSs is categorically pointed out. In the cases where NaSs are formed, there is no formation of horizons and the singularity is visible globally, as shown in Fig.(\ref{sub44}). The crucial difference between the two NaSs depicted here is that, the one in Subfig.(\ref{sub2}) is an event-like NaS, whereas the one in Subfig.(\ref{sub44}) is object-like \cite{JoshiBhattacharyya2025}. The PNaSs formed from a polynomial inflationary potential within the allowed parameter values (Subfig. \ref{param}) will be object-like and visible always, once they are formed. Therefore, these will be observationally consequential and could possibly shed light on a working theory of quantum gravity. 
Note that the ORNGs schematically depicted in Subfig.(\ref{sub44}) show that these singularities are visible to distant observers.
\subsection*{Exponential potential}
Another such example is a exponential potential \cite{Liddle:1998xm}:
\be \label{exp}
V(\phi) = V_0 \exp{(-\xi \phi)},
\ee
where, $V_0$ and $\lambda$ are scaling constants. It is very easy to note that for all values of these constants Eq.(\ref{Vprime}) is true. For physical scenarios, these scaling constants are non-zero. In such cases, Eq.(\ref{AH}) is given by:
\be 
\lim_{a\to 0} \exp{(-\xi\phi)} < \lim_{a\to 0} \frac{1}{a^2}.
\ee
From this, if we assume like before that $\phi \propto\frac{1}{a^p}$, then for $\xi>0$ the end-state of this collapse is a PBH. In contrary, if $\xi<0$, then the end-state is a globally naked object-like PNaS.
\subsection*{Double-well potential}
Double-well potentials are also suitable candidates for such a collapse. First proposed by Guth \cite{Guth:1980zm} in a cosmological context, this potential can be written as:
\be \label{DW}
V(\phi) = \mu \phi^2 + \lambda \phi^4,
\ee 
where, these constants determine the suitability of the inequality given in Eq.(\ref{Vprime}). From this, the allowed parametric regions can be found. From Fig.(\ref{fig5}), it can be seen that this allowed region for the threshold of overdensity, as given in Eq.(\ref{Vprime}), increases as the value of $\phi$ grows.}
\begin{figure}[ht!]
    \centering
    \begin{subfigure}[b]{0.4\textwidth}
        \centering
\includegraphics[width=\textwidth]{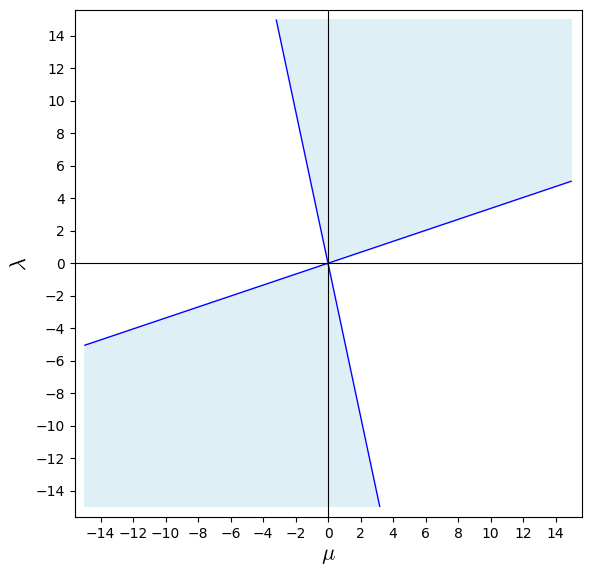}
    \caption{Parametric regions allowing inequality Eq.(\ref{Vprime}) for $\phi=0.85$.}
        \label{sub3}
    \end{subfigure}
    \hspace{25mm}
    \begin{subfigure}[b]{0.4\textwidth}
        \centering
        \includegraphics[width=\textwidth]{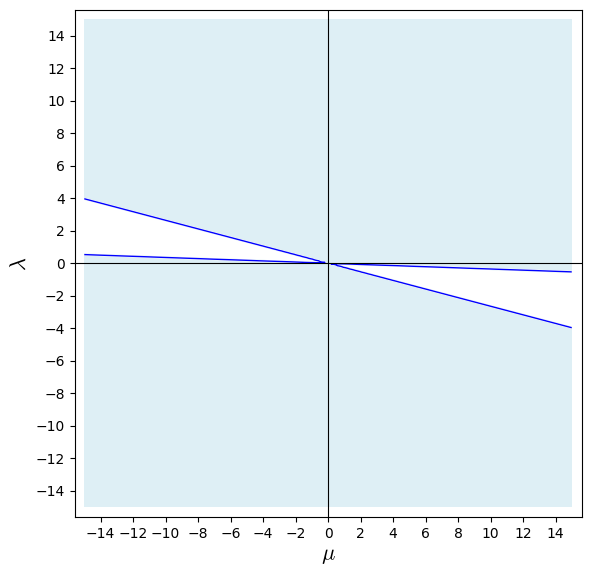} 
        \caption{FParametric regions allowing inequality Eq.(\ref{Vprime}) for $\phi=0.85$.}
        \label{sub4}
    \end{subfigure}
\caption{The inflationary potential $V(\phi)=\mu \phi^2 + \lambda \phi^4$ and the scale factor varying with the scalar field as $\phi(a)\propto 1/{a^p}$, it can be seen that the allowed parametric region increases as the collapse reaches its end-state.}
    \label{fig5}
\end{figure}
\textcolor{black}{If increased more, the whole region becomes allowed. As a specific instance, with $\mu = -\frac{16}{3}\lambda$ and employing Eq.(\ref{AH}), it can be seen that this kind of a potential ends up in a NaS as well. These parametric values are covered in the shaded regions of Fig.(\ref{fig5}). Thus, such potentials are suitable candidates for PNaSs as well.}

\textcolor{black}{\section{Collapsing bubble in an expanding universe}
We have discussed till now  the gravitational collapse, seeded by a scalar field and the visibility of its end-state. But it is crucial to note that this collapse occurs in a local patch: an overdense bubble. Outside this bubble, the rest of the universe is expanding, governed by flat FLRW geometry and we use Israel-Darmois matching conditions to ensure that these two processes can take place simultaneously without hampering the physical solution. For this, Israel-Darmois junction conditions are to be applied on the imbedded 3-hypersurface, on which these collapsing and expanding space-times lie side by side. The local co-ordinates of the overdense and background patches are identified by the suffixes $O$ and $B$ respectively. The induced line-element of the over-dense bubble is:
\be 
ds_O^2 = -dt^2_O + a_O^2 (t_O) [dr^2_O + r_b^2 d\Omega^2],
\ee 
and the background has the line element:
\be 
ds_B^2 = -dt^2_B + a_B^2 (t_B) [dr^2_B + r_b^2 d\Omega^2].
\ee
These two geometries have to match at a co-ordinate radius $r_b$ throughout the course of the collapse. So, the line-elements of the two-spheres of these two local geometries are the same. The first fundamental form makes the metric elements to match, which is the first criterion for a smooth matching and the following condition arises from this:
\be \label{match1}
r_b a_O (t_O) = r_b a_B (t_B). 
\ee 
To match the second fundamental form, the unit normal at the matching hypersurface has to be calculated first. For a flat FLRW metric, this unit normal is:
\be 
n^i = \left[0, \frac{1}{a(t)},0,0\right].
\ee 
The second fundamental form, extrinsic curvature is then calculated as:
\ba 
K_{ab}&=& \frac{1}{2} \mathcal{L}_n g_{ab} \nonumber\\
&=& \frac{1}{2}\left(g_{ab,c}n^c + g_{bc}n,_a^c+ g_{ac}n,_b^c\right), \label{ex}
\ea 
where, the extrinsic curvature is the Lie derivative of the metric tensor in the direction of the normal vector and Eq.(\ref{ex}) is the expansion of the same. Using this, the components of the extrinsic curvature can be found out. $K_{\theta\theta}$ components of the background and the bubble gives back the condition of Eq.(\ref{match1}). Similarly, the $K_{\tau\tau}$ component can be found out as:
\be \label{match2}
K_{\tau\tau}= \frac{F}{2r_b a} + r_b^2 a \ddot{a},
\ee
where, $F$ is the Misner-Sharp mass function. These quantities are functions of the local co-ordinates. Now, note that the background is expanding with positive acceleration, thus $\ddot{a}_B (t_B)>0$. Simultaneously, the bubble is collapsing and the acceleration must be decreasing, making $\ddot{a}_O (t_O)<0$. Therefore, $K^B_{\tau\tau}$ and $K^O_{\tau\tau}$ cannot be matched exactly and this needs the introduction of a thin-shell \cite{Goldwirth:1994ru, Mars:1993mj}. A simple question arises: if $a_{O} (t_O)=a_{B} (t_B)$, then how can $\ddot{a}_O(t_O)\neq \ddot{a}_B(t_B)$ be true throughout the collapse? This can be understood by differentiating $a(t(\tau))$ twice, with respect to proper time $\tau$. The acceleration can be found out to be:
\be 
\ddot{a} (t) = \frac{1}{\left(\frac{dt}{d\tau}\right)^2} \left[\ddot{a}(\tau) - \dot{a} \frac{d^2 t}{d\tau^2}\right].
\ee 
Here, $\ddot{a} (t)$ is the second order differential of the scale factor with respect to the co-ordinate time: $\ddot{a} (t)\equiv \frac{\partial^2 a(t)}{\partial t^2}$. But, this is different from the proper acceleration $\ddot{a}(\tau)\equiv \frac{\partial^2 a(t)}{\partial \tau^2}$. Now, it can be easily seen that $a_O (\tau)= a_B(\tau)$ does not mean $\ddot{a}_O\neq \ddot{a}_B$ necessarily. Rather the local lapse function $\frac{dt}{d\tau}$ can make the proper and co-ordinate acceleration different from each other. In summary, the co-ordinate time takes care of the mismatch between the acceleration of the background and the bubble.\\\\
Now, the bubble is seeded by a scalar field and for this the extrinsic curvature can be calculated from Eq.(\ref{match2}) as:
\be \label{match3}
K_{\tau\tau}^O = \frac{r_b^2 a_O^2}{2} \left[\frac{\dot{\phi}^2}{2} + V(\phi)\right] \left(1-\frac13 \phi,_{a_O}^2\right),
\ee 
which is obtained using Eq.(\ref{rho}), Eq.(\ref{ddota}) and Eq.(\ref{rhoabyrho}). Here, the overhead dots represent differentiation with respect to the local co-ordinates $t_O$. The same can also be obtained for the background:
\be \label{match4}
K^B_{\tau\tau}= \frac{F_B}{2r_b a_B} + r_b^2 a_B \ddot{a}_B,
\ee
where, $F_B$ is the Misner-Sharp mass of the background geometry and depends on the energy density of the same. This is kept as a free function and can change at different epochs of the Universe. Also, $\ddot{a}_B$ is differential of $a_B$ with respect to the local time co-ordinate $t_B$. It has already been discussed that the $K_{\tau\tau}$ of the two sides do not match and their difference is noted as the surface energy of the thin-shell between them:
\ba \label{match5}
P_\Sigma &=& \frac{1}{R_b^2} (K^O_{\tau\tau}-K^B_{\tau\tau})\nonumber\\
&=&\frac{1}{2} \left[\frac{\dot{\phi}^2}{2} + V(\phi)\right] \left(1-\frac13 \phi,_{a_O}^2\right)\nonumber\\
&-& \frac{F_B R_b}{2} + \frac{1}{a_B} \ddot{a}_B,
\ea
where, $R_b = a_B r_b = a_O r_b$, the physical radius of the boundary or the physical radius of the thin shell. For this solution to be physically viable, the surface pressure has to be positive and that is the condition arising from matching the second fundamental form. Thus, for this solution to be physically viable:
\be \label{match6}
\frac{1}{2} \left[\frac{\dot{\phi}^2}{2} + V(\phi)\right] \left(1-\frac13 \phi,_{a_O}^2\right)- \frac{F_B R_b}{2} + \frac{1}{a_B} \ddot{a}_B>0.
\ee 
The physical source of positive surface pressure can be multiple: cosmological phase transitions in bubbles \cite{Berezin:1987bc}, existence of gravitationally interacting domain walls \cite{Ipser:1983db}. These phenomena are well-studied in the early universe, and the specific physical phenomenon can be pinned down once the potential and the scalar fields are known. Another reason for this pressure can be core collapse of supernovae \cite{Muller:2020ard} and it is a well studied phenomena in high-energy astrophysics. This can point towards primordial singularities, produced later during the evolution of the universe.}
\section{Possible direction towards observing PNaSs}
The speculated observational signatures of PNaSs would be closely related to the mass of the specific model of PNaS under scrutiny, just like PBHs \cite{Carr:2025kdk}. But unlike PBHs, whose mass is the mass within its event horizon, the mass of PNaS is the Misner-Sharp mass. Hence it is flexible within a parameter range of the cut-off radius $r_b$, as mentioned in Eq.(\ref{AH}). Keeping this in mind, we now elaborate the possible ways to observe PNaS, in comparison to the well-posed techniques to observe PBHs \cite{Carr:2025kdk}. One notable way to detect primordial singularities is by observing the spectral distortion of the CMB spectrum due to injection of energy by the singularities \cite{Yang2022}. The injection of energy can be generated by two different processes: evaporation of the singularity or scattering by the singularity. These processes have been discussed for naked singularities \cite{Patil2012, Goswami2006}, and with modifications these could be used to constrain the masses of PNaSs using the observational data from COBE, FIRAS and PIXIE experiments \cite{Chluba:2025wxp}. Another way to detect singularities is to probe the lensing effects of them, which has been discussed
for PBHs \cite{Paczynski86} and a similar detection analysis can also be done for PNaSs \cite{Virbhadra:2002ju}.
The photons passing very close to a NaS could circle the NaS many more times than in the case for a BH before reaching the observer and hence the caustic for a NaS could be much brighter.
To observe PBHs, it is also prescribed to look at gravitational signatures \cite{Bird2016} captured by LIGO and LISA. Gravitational waves would also be emitted from PNaSs \cite{Malafarina:2016rdm} and these results need to be matched to the observations to either rule out or include PNaSs as dark matter candidates. The observational signatures of accretion disc luminosity of PBHs has been commented upon \cite{Gaggero2017} and a similar analysis can also be done for PNaSs \cite{Tahelyani:2022uxw, Chakraborty:2024jma}. 
There are multiply different ways to distinguish naked singularities from blackholes observationally \cite{Chakraborty:2022ltc, Chakraborty:2016mhx, Chakraborty:2016ipk, Chakraborty:2019rna}. Further detailed discussions on that can also be found in \cite{JoshiBhattacharyya2025} and the references within and these techniques can be used to observationally distinguish PNaSs frrom PBHs.

\section{Conclusion and discussions}

Gravitational collapse from primordial fluctuations in the space-time and subsequent formation of singularities in the early universe have been postulated some fifty years back and since then it has garnered a hefty load of curiosity and some validation through observational constraints. In this work, \textcolor{black}{we have studied the unhindered collapse, which is seeded by a scalar field with an arbitrary potential. Through this, we have pointed out that there are inflationary potentials whose dynamical collapse can lead up to either a PBH or a PNaS. We have also worked out some specific examples of such potentials which would lead up to the formation of a PNaS.} The current literature around primordial singularities has only discussed PBHs, but we have shown here that the dynamics of the collapse may or may not form an apparent horizon, which prohibits future causal connections between the singularity and any other events in space-time beyond this horizon. We have found out that the PNaSs formed in these scenarios are  object-like and visible at all future times after they are formed. Subsequently, we have separated the parametric regions for each of these cases: the one where an apparent horizon forms, leading to  PBHs, and if an apparent horizon does not form, leading to PNaSs. 
We have shown this for a largely general class of such scalar fields potentials. Our approach and length scales are coherent with the treatment in leading literature that discusses the formation of PBHs. We have also analysed the constraints on the slow-roll parameters of this potential that allows the abundance of the PNaSs formed to be high enough for detection.
These constraints, in no way, interfere with the absence of an AH in our analysis and we have shown it explicitly. \textcolor{black}{Furthermore, we have explicitly shown that such a collapse can occur in the background of a universe expanding at an accelerated rate, thus capturing the physical process of that very early epoch.} Thus, we have argued that PNaSs are likely to form in the early universe and general relativity presents no logical constraints against their formation. Possible mechanism for the formation of PNaSs has been elucidated in our treatment above.

Therefore, PNaSs could be ruled out only by observations.
However, one must proceed with utmost care to suggest observational constraints on PNaSs, as most of our scientific debates around observing the very early universe is largely speculative. 
The further we try to look back in time, more different are the physical situations from our currently observable universe. 
To compensate, we work with various symmetry assumptions and other theoretical priors, which in turn make the interpretation of observations speculative. 
Currently, this discussion is out of our scope of interest and we defer this to a future work. 

\section{Acknowledgments}
K and PSJ are grateful for the discussions and comments from Gaurav Goswami and Raghavan Rangarajan. The authors also thank Avi Loeb for his critical comments.

\end{document}